\documentclass[%
	floatfix,%
	reprint,%
	superscriptaddress,%
	nofootinbib,%
	twocolumn,%
	amsmath,%
	amssymb,%
	aps,%
	prapplied%
]{revtex4-2}

\usepackage{graphicx}
\usepackage{siunitx}
\usepackage{float}

\begin{document} %%%%%%%%%%%%%%%%%%%%%%%%%%%%%%%%%%%%%%%%%%%%%%%%%%%%%%%%%%%%%%
%%%%%%%%%%%%%%%%%%%%%%%%%%%%%%%%%%%%%%%%%%%%%%%%%%%%%%%%%%%%%%%%%%%%%%%%%%%%%%%

\title{Využití radioluminiscenčních zdrojů pro kalibraci \\ jednofotonových kamer}

\author{Radek Machulka}
	\email{radek.machulka@fzu.cz}
	\affiliation{
		Společná laboratoř optiky Univerzity Palackého a Fyzikálního ústavu Akademie věd České republiky, Přírodovědecká fakulta, Univerzita Palackého, Olomouc
	}

\author{Václav Michálek}
	\affiliation{
		Společná laboratoř optiky Univerzity Palackého a Fyzikálního ústavu Akademie věd České republiky, Přírodovědecká fakulta, Univerzita Palackého, Olomouc
	}

\author{Ondřej Haderka}
	\affiliation{
		Společná laboratoř optiky Univerzity Palackého a Fyzikálního ústavu Akademie věd České republiky, Přírodovědecká fakulta, Univerzita Palackého, Olomouc
	}

\author{Jan Peřina, Jr}
	\email{jan.perina.jr@upol.cz}
	\affiliation{
		Společná laboratoř optiky Univerzity Palackého a Fyzikálního ústavu Akademie věd České republiky, Přírodovědecká fakulta, Univerzita Palackého, Olomouc
	}

\begin{abstract}
V~tomto článku se věnujeme kalibraci kvantové účinnosti jednofotonových kamer s~využitím radioluminiscenčních světelných zdrojů. Tyto metody jsou následně porovnány s~absolutními kalibračními technikami založenými na detekci párových optických polí. V~závěru je pak navržena metoda transferu absolutní kalibrace právě pomocí zmíněných radioluminiscenčních zářičů.
\end{abstract}

\maketitle

%%%%%%%%%%%%%%%%%%%%%%%%%%%%%%%%%%%%%%%%%%%%%%%%%%%%%%%%%%%%%%%%%%%%%%%%%%%%%%%

\def\uv#1{,,#1``} % české uvozovky
\newcommand{\figocw}{\linewidth}
\newcommand{\reffig}[1]{\textit{obr.~\ref{#1}}}
\renewcommand{\fbox}{}

\section{Úvod}
Jednofotonové kamery, tj.~plošné detektory s~prostorovým rozlišením v~příčné rovině schopné detekovat individuální kvanta elektromagnetického pole, se stávají v~současné době stále častěji běžnými součástmi vybavení nejen optických laboratoří. Tyto kamery jsou charakteristické zejména nízkým šumem a vysokou detekční účinností. Díky těmto vlastnostem nacházejí uplatnění v~aplikacích zaměřených na detekci velmi slabých optických polí, jako je např. měření fotopulzních rozdělení~\cite{Haderka2005}.

Pro aplikace využívající čítání fotonů je však zcela nezbytná charakterizace použitého detektoru, zejména šumu, zisku a detekční účinnosti. V~tomto článku se budeme věnovat primárně poslední veličině, tj. určení detekční (kvantové) účinnosti.

\section{Kalibrace pomocí radioluminiscenčních světelných zdrojů}
Běžný způsob kalibrace světelných detektorů spočívá v~použití kalibrovaných zdrojů osvětlení, nejčastěji wolframových halogenových lamp~\cite{Hollandt2005}. Použití těchto zdrojů s~sebou nese ovšem celou řadu omezení, jako jsou např. nízká životnost, potřeba stabilních kalibrovaných zdrojů napětí a zejména pak nutnost časté rekalibrace. Pro mnoho pracovišť je tedy snaha o~kalibraci detekčních aparatur příliš komplikovaná a nákladná a ve výsledku na ni pak často rezignují.

Zajímavou alternativu oproti těmto standardním termálním zdrojům proto představují zdroje radioluminiscenční. Jedná se o~světelné zářiče využívající luminiscence způsobené excitací atomů ionizujícím zářením generovaným rozpadem některého radionuklidu. Typickým příkladem takového zářiče je tzv. GTLS zdroj (gaseous tritium light source), který využívá tritia, ra\-dio\-ak\-tiv\-ní\-ho izotopu vodíku v~plynné formě. Tento slabý a relativně bezpečný beta zářič, jehož poločas rozpadu je zhruba 12,32~let, pak indukuje luminiscenci v~luminoforu (nejčastěji sulfid zinečnatý, ZnS) naneseném na vnitřní straně kapsle z~borosilikátového skla, v~níž je uzavřen.

Dlouhodobá stabilita fotonového toku (daná ve střední hodnotě predikovatelným radioaktivním rozpadem), relativně dlouhá doba života, kompaktní rozměry, absence napájení a v~neposlední řadě i nízké pořizovací a provozní náklady pak předurčují použití takových zdrojů jako radiometrických standardů.

Velice podrobný popis GTLS zářiče, včetně jeho použití pro kalibraci obrazových senzorů, je uveden v~článku \cite{McFadden2022}, kde se autoři věnují nejen charakterizaci vlastního zdroje, ale i návrhu konkrétního kalibračního uspořádání včetně vlastní kalibrace. V~části věnující se kalibraci kamer se nicméně zabývají primárně kalibrací zisku a temných detekčních událostí, zatímco o~kalibraci detekční účinnosti se zmiňují pouze okrajově, jelikož ta vyžaduje počáteční kalibraci zářiče. Tu je pochopitelně možné provést standardním způsobem, jakým se kalibrují již zmíněné termální zdroje, nicméně existuje mnohem sofistikovanější a přesnější způsob založený na detekci párových optických polí.

\section{Kalibrace využívající detekce párových polí}
Párovými poli rozumíme optická pole sestávající pouze z~jednotlivých fotonových párů. Typickým zdrojem takových polí je parametrický proces sestupné frekvenční konverze (SPDC), kde dochází ve vhodném nelineárním prostředí k~rozpadu čerpacího fotonu na fotonový pár \cite{Burnham1970}. Takový proces musí přirozeně vyhovovat zákonům zachování, které pak předurčují frekvence a směry šíření generovaných fotonů. V~důsledku vlastností tohoto procesu vykazují jednotlivé fotony z~každého páru, velmi silné neklasické korelace, tzv. entanglement. Pro potřeby radiometrické kalibrace je zásadní fakt, že fotony jsou generovány vždy současně, tj. v~párech. Historicky se jednotlivé fotony tvořící fotonový pár (svazky, jimiž se šíří) nazývají signálový a jalový.

Vlastní kalibrace pomocí fotonových párů je založena na jednoduchém principu uvedeném v~\cite{Migdall1999}. Detekuje-li detektor foton v~signálovém svazku, je přítomen foton i v~jalovém svazku. Jalový detektor je tedy v~tomto případě kalibrovaný s~jednotkovou pravděpodobností. Pro velký počet opakování pak můžeme definovat vztah pro kvantovou účinnost detektoru ve tvaru $\eta_{\rm i,s} = \langle m_{\rm s}m_{\rm i} \rangle / \langle m_{\rm s,i} \rangle$, kde výraz v~čitateli představuje počet tzv. koincidencí, tj. současných detekčních událostí v~obou svazcích, zatímco výraz ve jmenovateli je střední počet detekčních událostí v~jednom svazku. Indexy i, s pak popisují jednotlivé svazky, i je přiřazen jalovému (z~anglického idler) svazku, zatímco s představuje signálový svazek.

Předchozí přístup je však zřejmě platný pouze v~ideálním případě, kdy detekované optické pole je skutečně striktně párové. V~reálné situaci se ovšem střetáváme se skutečností, kdy je v~obou svazcích přítomna jistá úroveň jednofotonového šumu. Avšak tento případ vyžaduje mnohem sofistikovanější přístup založený na měření sdruženého fotopulzního rozdělení \cite{PerinaJr2013a}. Toto experimentální rozdělení je pak aproximováno vhodným teoretickým modelem. Výsledkem takového teoretického proložení jsou nejen parametry jednotlivých párových a šumových polí, ale rovněž i hledané detekční účinnosti. Kompletní popis takové kalibrační techniky je uveden v~\cite{PerinaJr2012a,PerinaJr2014}.

Předchozí kalibrační technika, ačkoli poskytuje velmi přesnou absolutní kalibraci detekční účinnosti, představuje značnou obtíž zejména z~experimentálního pohledu. Prvně je třeba realizovat vhodný zdroj fotonových párů, následně je vyžadováno značné množství měření, jelikož měřený histogram musí být pro potřeby následného zpracování získán s~dostatečně malými nejistotami. Taková měření mohou probíhat v~řádu hodin až dní, což přináší další výzvy zejména z~pohledu stability celého uspořádání. Získaná experimentální data je pak třeba aproximovat definovaným teoretickým modelem, což zpravidla vyžaduje použití výpočetně náročných iterativních algoritmů využívajících metody maximální věrohodnosti \cite{PerinaJr2012}.

\section{Transfer absolutní kalibrace}
Z~předchozích odstavců tedy vyplývá, že optimální kalibrační strategie by mohla být založena na kombinaci obou zmíněných metod. Robustní absolutní kalibrace založená na rekonstrukci párových polí bude použita pro počáteční kalibraci fotonového toku radioluminiscenčního zářiče, který bude možné posléze využít pro kalibraci dalších detektorů bez nutnosti provádět dlouhotrvající komplikovaná měření. Takto kalibrovaný zdroj je navíc možné velmi snadno sdílet mezi laboratořemi, což dále snižuje celkové náklady na experimentální vybavení jednotlivých pracovišť. Tento přístup navíc umožňuje kalibraci detektorů, které z~různých důvodů absolutní kalibraci pomocí fotonových párů nedovolují, např. neposkytují rozlišení v~počtu fotonů apod.

\begin{figure}[H]
	\centering
	\fbox{
		\includegraphics[width=\figocw]{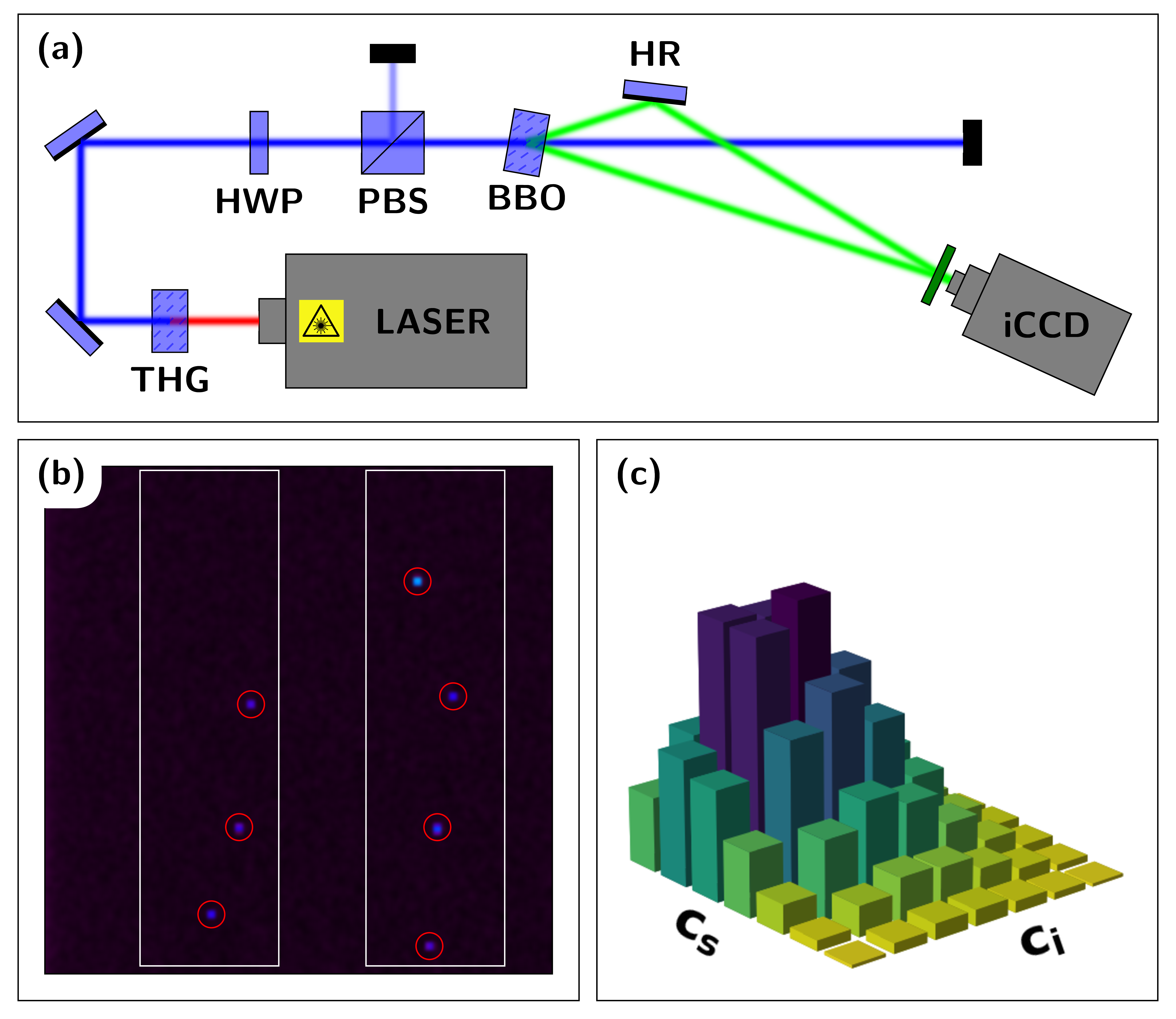}
	}
	\caption{Schéma experimentálního uspořádání pro měření fotopulzního rozdělení párových polí~(a) včetně typického snímku z~kamery obsahujícího jednotlivé detekční události~(b) a výsledného sdruženého 2D histogramu získaného z~celkového počtu~1,24 $ \times 10^6 $~snímků. THG:~generátor třetí harmonické, HWP:~půlvlnná destička, PBS:~polarizátor, BBO:~$\beta$-BaB$_{2}$O$_{4}$ nelineární krystal, iCCD:~intenzifikovaná CCD kamera, $c_{\rm s,i}$:~počty signálových, resp.~jalových fotoelektronů.}
	\label{fig:01_setup}
\end{figure}

Pro praktickou demonstraci zmíněného postupu lze využít data získaná z~experimentálního uspořádání jehož schéma je uvedeno v~\reffig{fig:01_setup}. Podrobný popis experimentu je pak uveden např. v~\cite{Haderka2005}.

Čerpací svazek, získaný z generátoru třetí harmonické frekvence využívající pole fundamentální frekvence femtosekundového laseru, je výkonově stabilizován pomocí kombinace půlvlnné destičky a polarizátoru, načež čerpá parametrický proces v~nelineárním krystalu. Část pole generovaného v~kuželové ploše je detekována kamerou přímo, zatímco konjugovaná část po odraze na zrcadle. V~tomto konkrétním případě je detektorem intenzifikovaná CCD (iCCD) kamera. Spektrální a prostorová šířka detekovaných svazků je pak omezena pomocí interferenčního filtru. Signální a jalový svazek jsou tedy detekovány iCCD kamerou v~různých částech fotokatody. Tímto způsobem je možné experimentálně generovat histogram $f(c_{\rm s}, c_{\rm i})$ reprezentující pravděpodobnost současné detekce~$c_{\rm s}$ a $c_{\rm i}$ fotoelektronů v~signálovém a jalovém svazku. Výsledné detekční účinnosti $\eta_{\rm s}$, $\eta_{\rm i}$ pak získáme minimalizací odchylek teoretického fotopulzního rozdělení $p_{c}(c_{\rm s}, c_{\rm i})$ a experimentálního histogramu~$f(c_{\rm s}, c_{\rm i})$:
\begin{equation}
	\mathcal{D} = \sqrt{\sum_{c_{\rm s},c_{\rm i} = 0}^{\infty}%
		\left[ p_{c}(c_{\rm s}, c_{\rm i}) - f(c_{\rm s}, c_{\rm i}) \right]^{2}}.
\end{equation}
Podrobný popis teoretického modelu vedoucího k~veličině $p_{c}$ je podrobně uveden např. v~\cite{Haderka2005}.

Hodnoty takto získaných detekčních účinností v~tomto konkrétním případě činí 0,243$\pm 0,05$ pro signální, resp. 0,235$ \pm 0,05$ pro jalový svazek. Rozdílné hodnoty jsou pak důsledkem celkové geometrie experimentu, kdy získaná detekční účinnost představuje celkovou pravděpodobnost detekce fotonu včetně propagace experimentálním uspořádáním.

\begin{figure}[H]
	\centering
	\fbox{
		\includegraphics[width=\figocw]{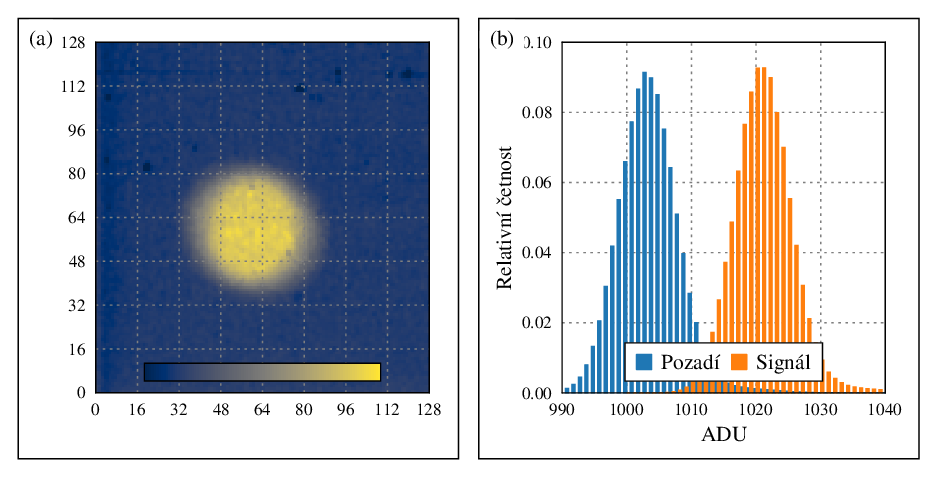}
	}
	\caption{Typický akumulovaný snímek z~iCCD kamery osvětlené GTLS zdrojem (a) včetně výsledného histogramu šumových a signálových snímků~(b).}
	\label{fig:02_source}
\end{figure}

S~takto kalibrovaným detektorem je pak možné kalibrovat vlastní radioluminiscenční zářič a ten pak použít jako radiometrický standard. Typický akumulovaný snímek z~iCCD kamery osvícené tímto zářičem je uveden v~\reffig{fig:02_source} včetně celkového histogramu z~CCD čipu kamery.

Analýza obrazových dat probíhá identicky jako v~případě detekce párových polí, nyní již však bez rozlišení na signálový a jalový svazek. Zpracování hrubých dat je pak závislé na konkrétní konstrukci kamery. Pro iCCD kamery probíhá typicky v~několika krocích. Od každého signálového snímku se nejprve odečte průměrný snímek pozadí, čímž se měření opraví na případné nehomogenity v~odezvě senzoru kamery. Takto opravené snímky se následně \uv{zaprahují}, kdy hodnota prahu reflektuje šířku rozdělení šumu pozadí. Výsledná kvantová účinnost je tedy ve výsledku dána kombinací jak samotného zesílení kamery, tak i jejích šumových vlastností. V~\uv{zaprahovaných} snímcích pak lze identifikovat vlastní detekční události, které jsou v~důsledku konstrukce iCCD kamer často rozprostřené přes několik sousedních pixelů (viz \reffig{fig:01_setup}{\it b}). Z~celé série měření pak získáme střední počet detekčních událostí, které pak můžeme snadno porovnat s~očekávanou intenzitou zářiče a získat tak vlastní detekční účinnost kamery.

\begin{figure}[H]
	\centering
	\fbox{
		\includegraphics[width=\figocw]{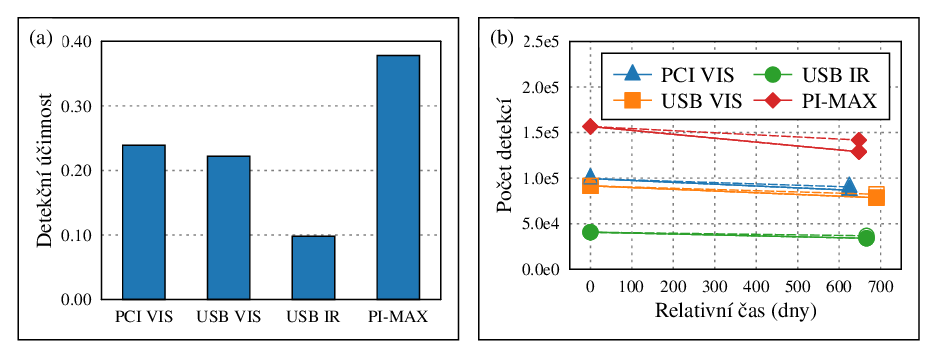}
	}
	\caption{Porovnání detekčních účinností různých jednofotonových kamer~(a) včetně poklesu intenzity zářiče za delší časový úsek~(b). Plná čára představuje výsledky měření, čárkovaná čára očekávaný pokles odpovídající poločasu rozpadu tritia.}
	\label{fig:03_detections}
\end{figure}

Výsledné detekční události, získané transferem absolutní kalibrace pomocí párových polí, jsou uvedeny v~grafu v~\reffig{fig:03_detections}. Celkově byly použity čtyři různé iCCD kamery: Andor DH734-18F-63-9FO, kamera s~externí digitalizační jednotkou~(PCI-VIS), Andor DH334T-18U-63, alternativní verze předchozí kamery s~integrovaným digitizérem~(USB-VIS), Andor DH334T-18U-73, identická verze s~katodou intenzifikátoru optimalizovanou pro infračervenou oblast~(USB-IR) a Princeton Instruments PI-MAX PM4-1024f-HB-FG-18-P43-CM, iCCD kamera konkurenčního výrobce se sníženým vyčítacím šumem (PI-MAX). Z~grafu jsou pak patrné detekční účinnosti jednotlivých kamer. Minimální účinnosti podle očekávání dosahuje kamera optimalizovaná pro infračervenou oblast, maximální pak kamera se sníženým vyčítacím šumem, která umožňuje nižší úroveň \uv{prahování}. Obě verze kamer Andor pak dosahují podobných hodnot účinnosti s~malým rozdílem ve prospěch PCI verze dané opět nižším vyčítacím šumem externího digitizéru.

V~\reffig{fig:03_detections} je současně uveden graf zobrazující pokles intenzity zářiče v~důsledku radioaktivního rozpadu. Z~grafu nicméně vyplývá, že pokles získaný měřením je vyšší než pokles předpokládaný čistě na základě vlastního rozpadu. Existuje důvod se domnívat, že tento systematický pokles je primárně způsoben degradací luminoforu a představuje hlavní omezení pro potřeby rekalibrace zářiče~\cite{Swart2004}.

\section{Závěr}
Kalibrace obrazových senzorů vykazující jednofotonovou citlivost je komplikovaná záležitost primárně z~toho důvodu, že vyžaduje kalibrovaný zdroj osvětlení s~velmi nízkou intenzitou. Existuje však robustní alternativa v~podobě absolutní kalibrace pomocí párových optických polí. Tato metoda ovšem vyžaduje komplikované experimentální uspořádání, velmi dlouhé doby měření a v~neposlední řadě i výpočetně náročnou rekonstrukci detekovaného pole. To omezuje použití těchto pokročilých metod na specializovaná optická pracoviště.

Jako kompromisní řešení se nabízí transfer absolutní kalibrace pomocí radioluminiscenčních zářičů. Výhoda takového řešení spočívá v~možnosti velmi přesné kalibrace celé řady dalších detektorů, či jejich provozních režimů, bez nutnosti komplikovaných měření, jakož i možnost sdílení lehce přenosného zdroje mezi laboratořemi. Hlavní omezení při použití radioluminiscenčních zářičů pak představuje degradace luminoforu, která způsobuje rychlejší pokles intenzity zářiče vůči očekávanému poklesu způsobenému vlastním radioaktivním rozpadem. Tento efekt je třeba zohlednit při stanovení vhodného intervalu rekalibrace zářiče.

Na úplný závěr dodejme, že v~poslední době směřuje značná pozornost ke zpracování obrazových dat z~kamer prostřednictvím metod strojového učení. Vhodně navržené modely dokáží velmi účinně parametrizovat celou řadu reálných úloh, jako jsou zde zmíněné rekonstrukce detekovaných polí, včetně odhadu konkrétních parametrů (např. právě detekčních účinností), až po stanovení komplexních metrik optických polí (např. míry neklasičnosti), jejichž kvantifikace je značně obtížná a výpočetně náročná~\cite{Machulka2024}. Přestože použití těchto metod přináší netriviální míru složitosti, zejména s~ohledem na počáteční návrh modelu a jeho následný trénink, metody umožňují získat věrnou charakterizaci zkoumaných systémů.

%%%%%%%%%%%%%%%%%%%%%%%%%%%%%%%%%%%%%%%%%%%%%%%%%%%%%%%%%%%%%%%%%%%%%%%%%%%%%%%

% \vspace{3mm} \noindent \textbf{Data Availability.}
% Data underlying the results presented in this work may be obtained from <xxx> repository <xxx>.

\vspace{3mm} \noindent
\textbf{Autoři děkují projektu ITI CZ.02.01.01/00/23\_021/0008790 Ministerstva školství, mládeže a tělovýchovy České republiky a Evropské unii.}

\vspace{3mm} \noindent
\textbf{Článek byl publikován pod licencí CC BY 4.0.}

%%%%%%%%%%%%%%%%%%%%%%%%%%%%%%%%%%%%%%%%%%%%%%%%%%%%%%%%%%%%%%%%%%%%%%%%%%%%%%%

\bibliography{machulka}

%%%%%%%%%%%%%%%%%%%%%%%%%%%%%%%%%%%%%%%%%%%%%%%%%%%%%%%%%%%%%%%%%%%%%%%%%%%%%%%
\end{document}